\documentclass[twocolumn]{aastex631}

\let\tablenum\relax
\usepackage{hyperref}

\usepackage[separate-uncertainty,multi-part-units = single,group-digits=false]{siunitx}
\usepackage{bm}

\usepackage{natbib}

\usepackage{chemformula}
\usepackage{amsmath}
\usepackage{float}

\setcitestyle{notesep={ }}
\shorttitle{Discovery of AF Lep b}
\shortauthors{Franson et al.}

\graphicspath{{./}{figures/}}

\begin{document}

\title{Astrometric Accelerations as Dynamical Beacons: \\ A Giant Planet Imaged Inside the Debris Disk of the Young Star AF Lep}

\author[0000-0003-4557-414X]{Kyle Franson}
\altaffiliation{NSF Graduate Research Fellow} \affiliation{Department of Astronomy, The University of Texas at Austin, Austin, TX 78712, USA}

\author[0000-0003-2649-2288]{Brendan P. Bowler}
\affiliation{Department of Astronomy, The University of Texas at Austin, Austin, TX 78712, USA}

\author[0000-0003-2969-6040]{Yifan Zhou}
\altaffiliation{51 Pegasi b Fellow}
\affiliation{Department of Astronomy, The University of Texas at Austin, Austin, TX 78712, USA}

\author{Tim D. Pearce}
\affiliation{Astrophysikalisches Institut und Universit\"{a}tssternwarte, Friedrich-Schiller-Universit\"{a}t Jena, Schillerg\"{a}{\ss}chen 2–3, 07745 Jena, Germany}

\author[0000-0001-8170-7072]{Daniella C. Bardalez Gagliuffi}
\affiliation{Department of Physics \& Astronomy, Amherst College, 25 East Drive, Amherst, MA 01003, USA}
\affiliation{American Museum of Natural History, 200 Central Park West, New York, NY 10024, USA}

\author{Lauren Biddle}
\affiliation{Department of Astronomy, The University of Texas at Austin, Austin, TX 78712, USA}

\author{Timothy D. Brandt}
\affiliation{Department of Physics, University of California, Santa Barbara, Santa Barbara, CA 93106, USA}

\author[0000-0003-0800-0593]{Justin R. Crepp}
\affiliation{Department of Physics, University of Notre Dame, 225 Nieuwland Science Hall, Notre Dame, IN 46556, USA}

\author[0000-0001-9823-1445]{Trent J. Dupuy}
\affiliation{Institute for Astronomy, University of Edinburgh, Royal Observatory, Blackford Hill, Edinburgh, EH9 3HJ, UK}

\author[0000-0001-6251-0573]{Jacqueline Faherty}
\affiliation{American Museum of Natural History, 200 Central Park West, New York, NY 10024, USA}

\author[0000-0003-0054-2953]{Rebecca Jensen-Clem}
\affiliation{Astronomy \& Astrophysics Department, University of California, Santa Cruz, CA 95064, USA}

\author[0000-0003-4022-6234]{Marvin Morgan}
\affiliation{Department of Astronomy, The University of Texas at Austin, Austin, TX 78712, USA}

\author[0000-0002-1838-4757]{Aniket Sanghi}
\affiliation{Department of Astronomy, The University of Texas at Austin, Austin, TX 78712, USA}

\author[0000-0002-9807-5435]{Christopher A. Theissen}
\altaffiliation{NASA Sagan Fellow}
\affiliation{Center for Astrophysics and Space Sciences, University of California, San Diego, 9500 Gilman Drive, La Jolla, CA 92093, USA}

\author[0000-0001-6532-6755]{Quang H. Tran}
\affiliation{Department of Astronomy, The University of Texas at Austin, Austin, TX 78712, USA}

\author[0000-0002-1406-8829]{Trevor N. Wolf}
\affiliation{Department of Aerospace Engineering and Engineering Mechanics, The University of Texas at Austin, Austin, TX 78712, USA}
 
\begin{abstract}
We present the direct imaging discovery of a giant planet orbiting the young star AF Lep, a 1.2~$M_{\odot}$ member of the 24 $\pm$ 3 Myr $\beta$ Pic moving group.  AF Lep was observed as part of our ongoing high-contrast imaging program targeting stars with astrometric accelerations between Hipparcos and Gaia that indicate the presence of substellar companions. Keck/NIRC2 observations in $L'$ with the Vector Vortex Coronagraph reveal a point source, AF Lep b, at ${\approx}340$~mas which exhibits orbital motion at the 6-$\sigma$ level over the course of 13 months.  
A joint orbit fit yields precise constraints on the planet's dynamical mass of 3.2$^{+0.7}_{-0.6}$~$M_\mathrm{Jup}$, semi-major axis of 8.4$^{+1.1}_{-1.3}$~au, and eccentricity of 0.24$^{+0.27}_{-0.15}$.
AF Lep hosts a debris disk located at $\sim$50~au, but it is unlikely to be sculpted by AF Lep b, implying there may be additional planets in the system at wider separations. The stellar inclination ($i_*$ = 54$^{+11}_{-9}\degr$) and orbital inclination ($i_o$ = 50$^{+9}_{-12}\degr$) are in good agreement, which is consistent with the system having spin-orbit alignment.
AF Lep b is the lowest-mass imaged planet with a dynamical mass measurement and highlights the promise of using astrometric accelerations as a tool to find and characterize long-period planets.

\end{abstract}

\keywords{Direct imaging (387) --- Extrasolar gaseous giant planets (509) --- Orbit determination (1175) --- Debris disks (363) --- Astrometric exoplanet detection (2130)}

\section{Introduction \label{sec:intro}}
A growing number of planets with masses spanning 1--13 $M_\mathrm{Jup}$ and separations from $\approx$5 to 100~au have been imaged over the past 15 years (e.g., \citealt{Marois:2008ei}; \citealt{Marois:2010gp}; \citealt{Lagrange:2010fs}; \citealt{Rameau:2013vh}; \citealt{Macintosh:2015fw}; \citealt{Chauvin:2017aa}; \citealt{Keppler:2018dd}).  
It remains unclear whether these long-period ``super Jupiters'' predominantly form through a bottom-up core accretion scenario \citep{Pollack:1996jp}, possibly assisted by pebble accretion \citep{Johansen:2017im}; through a top-down gravitational instability route \citep{Boss:1997di}; or even via dynamical capture in a dense cluster at an early age \citep{Perets:2012cv}. Further complicating the picture, the star formation process is expected to produce objects with masses as low as the opacity limit of ${\sim}\SI{3}{M_{Jup}}$ through cloud fragmentation \citep{Bate:2009br}. Demographic studies are hinting that imaged wide-separation planets have a bottom-heavy mass distribution (\citealt{Wagner:2019iy}), a semi-major axis distribution weighted towards smaller separations (\citealt{Nielsen:2019cb}), an eccentricity distribution peaked at low values in contrast to higher-mass substellar companions (\citealt{Bowler:2020hk}; \citealt{nagpal:2023aa}), and preferentially aligned stellar obliquities \citep{Bowler2023}. These results resemble the demographics of closer-in planets found with radial velocities (e.g., \citealt{Fulton:2021cn}), although the significance of this similarity is limited by the small number of imaged planets.

An efficient strategy to identify long-period companions is to use the gravitational reflex motion they induce on their host stars.  Beginning with the discovery of the brown dwarf companion HR 7672 B (\citealt{Liu:2002fx}),
radial velocity trends have been repeatedly used to identify promising targets for follow-up high-contrast imaging (e.g., \citealt{Crepp:2014ce}; \citealt{Cheetham:2018ha}; \citealt{Bowler:2021ir}).
More recently, astrometric accelerations using proper motion differences between Hipparcos and Gaia
have proven to be similarly advantageous as a strategy to efficiently search for and characterize
substellar companions (e.g., \citealt{DeRosa:2019aa}; \citealt{Brandt:2019ey}; \citealt{Franson:2022bl}; \citealt{Hinkley:2022aa}; \citealt{currieDirectImagingAstrometric_2023}).

The Astrometric Accelerations as Dynamical Beacons program is a high-contrast imaging survey targeting a 
dynamically informed sample of young accelerating stars identified
in the Hipparcos-Gaia Catalog of Accelerations (HGCA; \citealt{Brandt:2021cd}).
Our targets are selected for youth, proximity, and from individual predictions of the
companion mass and separation required to produce the observed proper motion difference
between Hipparcos and Gaia EDR3. The first discovery from this program was the low-mass brown dwarf HIP 21152 B (\citealt{Franson:2023bb}), 
a T-dwarf companion in the Hyades cluster.

Here we present the discovery of a giant planet orbiting the young star AF Lep\footnote{AF Lep b was also independently and contemporaneously discovered by discovered by \citet{derosaDirectImagingDiscovery_2023} and \citet{mesaAfLepLowest_2023}, both using VLT/SPHERE.} based on observations with the NIRC2 Vector Vortex Coronagraph at the Keck II telescope. In Section~\ref{sec:aflep}, we summarize the physical and astrometric properties of AF Lep.  Our NIRC2 observations and analysis of the AF Lep light curve from the Transiting Exoplanet Survey Satellite (TESS) are described in Section~\ref{sec:observations}.  In Section~\ref{sec:results}, 
the relative astrometry of AF Lep b is used to confirm common proper motion with
its host star and carry out an orbit fit.  We also compare the orbital inclination
with the stellar inclination to assess consistency with spin-orbit alignment.
A summary of our results can be found in Section~\ref{sec:summary}.

\begin{deluxetable}{lcc}
\tabletypesize{\footnotesize}
\tablecaption{\label{tab:prop}Properties of AF Lep A and b}
\tablehead{\colhead{Property} & \colhead{Value} & \colhead{Refs}}
\startdata
\multicolumn{3}{c}{AF Lep A} \\
\hline
$\alpha_{2000.0}$ & 05:27:04.76 & 1\\
$\delta_{2000.0}$ & --11:54:03.5 & 1\\
$\mu_{\alpha}$\tablenotemark{a} (mas\,$\mathrm{yr^{-1}}$) & 16.915 $\pm$ 0.018 & 1 \\
$\mu_{\delta}$ (mas\,$\mathrm{yr^{-1}}$) & --49.318 $\pm$ 0.016  & 1 \\
$a$\tablenotemark{b} ($\si{mas.yr^{-2}}$) & $0.0191 \pm 0.0024$ & 3\\
$a$\tablenotemark{b} ($\si{m.s^{-1}yr^{-1}}$) & $2.43 \pm 0.30$ & 3\\
$\pi$ (mas) & $37.254 \pm 0.020$ & 1\\
Distance (pc) & $26.843 \pm 0.014$ & 1\\
SpT & F8V & 2\\
Mass ($M_\odot$) & $1.20 \pm 0.06$ & 3\\
Radius ($R_{\odot}$) & 1.25 $\pm$ 0.06 &   4 \\
Age (Myr) & 24 $\pm$ 3 & 5\\
$T_{\mathrm{eff}}$ (K) & 6130 $\pm$ 60 & 6\\
$\mathrm{[Fe/H]}$ (dex) & ${+}0.19$ $\pm$ 0.02 & 7 \\
$v \sin i$ ($\mathrm{km/s}$) & $50 \pm 5$ & 8 \\
$\log(R'_{HK})$ (dex) & --4.27 & 9 \\
$\log(L_X/L_{\mathrm{bol}})$ (dex) & --3.48 & 10 \\
$P_\mathrm{rot}$ (d) & 1.007 $\pm$ 0.009 & 11 \\
$i_*$ ($\degr$)  & 54$^{+11}_{-9}$  &  11 \\
$\mathrm{RUWE_{DR3}}$ & 0.918 & 1\\
$B$ (mag) & $6.83 \pm 0.02$ & 12\\
$V$ (mag) & $6.30 \pm 0.01$ & 12\\
Gaia $G$ (mag) & $6.210 \pm 0.003$ & 1\\
$J$ (mag) & $5.27 \pm 0.03$ & 13\\
$H$ (mag) & $5.09 \pm 0.03$ & 13\\
$K_s$ (mag) & $4.93 \pm 0.02$ & 13\\
$W1$ (mag) & $4.92 \pm 0.07$ & 14 \\
\hline
\multicolumn{3}{c}{AF Lep b} \\
\hline
$\rho_{2021.970}$ (mas) & 338 $\pm$ 11 & 11 \\
$\theta_{2021.970}$ ($\degr$) & 62.8 $\pm$ 1.3 & 11 \\
$\rho_{2023.090}$ (mas) & 342 $\pm$ 8 & 11 \\
$\theta_{2023.090}$ ($\degr$) & 72.0 $\pm$ 1.0 & 11 \\
$\Delta L'$ (mag) & 9.94 $\pm$ 0.14 & 11\\
$L'$ (mag) & 14.87 $\pm$ 0.15 & 11\\
$M_{L'}$ (mag) & 12.72 $\pm$ 0.15 & 11\\
$\log(L/L_{\mathrm{bol}})$ (dex)& $-4.81 \pm 0.13$ & \\
Dynamical Mass ($M_\mathrm{Jup}$) & 3.2$^{+0.7}_{-0.6}$ & 11\\
Semi-major Axis (au) & $8.4^{+1.1}_{-1.3}$ & 11\\
$e$ & $0.24^{+0.27}_{-0.14}$ & 11\\
$i_o$ (\degr) & $50^{+9}_{-12}$ & 11\\
Period (yr) & $22 \pm 5$ & 11\\
\enddata
\tablenotetext{a}{Proper motion in R.A. includes a factor of $\cos \delta$.}
\tablenotetext{b}{Calculated from proper motion difference between Hipparcos-Gaia joint proper motion and Gaia EDR3 proper motion in \citet{Brandt:2021cd}.}
\tablerefs{
(1) \citet{GaiaCollaboration:2022aa}; 
(2) \citet{Gray:2006aj}; 
(3) \citet{Kervella:2022aa}; 
(4) \citet{Kervella:2004aa};
(5) \citet{Bell:2015mnras}; 
(6) \citet{Ammons:2006uh}; 
(7) \citet{Gaspar:2016aa}; 
(8) \citet{Glebocki:2005aa}; 
(9) \citet{Isaacson:2010gk}; 
(10) \citet{Wright:2011dj}; 
(11) This work; 
(12) \citet{Hog:2000wk}; 
(13) \citet{Skrutskie:2006hl}; 
(14) \citet{cutriVizierOnlineData_2012}
}
\end{deluxetable}

\section{AF Lep: A Young Sun-Like Star \label{sec:aflep}}
AF Leporis ($=$AF Lep, HD~35850, HIP~25486, HR~1817) is a bright ($V = \SI{6.30 \pm 0.01}{mag}$; \citealt{Hog:2000wk}) F8V star \citep{Gray:2006aj} in the $\beta$ Pic moving group \citep{Zuckerman:2004araa}. It has a distance of 26.84\,pc \citep{GaiaCollab:2021aa}, an age of $24\pm3$\,Myr \citep{Bell:2015mnras}, and a mass of $1.20 \pm 0.06 \, M_\mathrm{\odot}$ \citep{Kervella:2022aa}. An infrared excess based on WISE, Spitzer/MIPS, and Herschel/PACS photometry indicates that AF Lep has a debris disk, and spectral energy distribution (SED) modeling implies the disk radius is located at $46 \pm 9$ au, although its unresolved nature means it could conceivably lie anywhere from 30 to 70 au \citep{Pawellek:2021mnras,Pearce:2022aa}. Using dynamical arguments, \citet{Pearce:2022aa} inferred that a single planet with a minimum mass of $1.1\pm0.2\,M_\mathrm{Jup}$ at a maximum semi-major axis of $35\pm6$~au would truncate the disk's inner edge. The minimum planetary mass increases steeply as its semi-major axis decreases, so a more massive planet closer in could also interact with the disk in a similar fashion. In a multi-planet scenario, the required planet mass is much lower ($0.09\pm0.03\,M_\mathrm{Jup}$). AF~Lep has been frequently targeted in direct-imaging exoplanet searches \citep{Biller:2013apj,Stone:2018aj,Nielsen:2019cb,Launhardt:2020aa}. These searches did not find any close companions.

AF Lep is listed as an SB2 in the chromospherically active binary catalog complied by \citet{Eker:2008mnras}. The initial double-lined spectroscopic binary classification was reported in \citet{Nordstrom:2004aa}, who found broad line profiles with structure and variable central positions (B. Nordstr\"om 2022, priv. communication). This variability may be caused by a close binary, but could also originate from strong imprints from starspot-related activity. Indeed, \citet{ZunigaFernandez:2021if} did not identify AF Lep as a single- or double-lined spectroscopic binary based on their recent high-resolution optical spectroscopy of this system. Additional spectroscopic monitoring would help clarify the binary nature of the host star. Because the original classification of AF Lep as an SB2 has not been validated with follow-up observations, for this work we treat AF Lep as a single active young star.

\begin{figure*}
    \centering
    \includegraphics[width=1\textwidth]{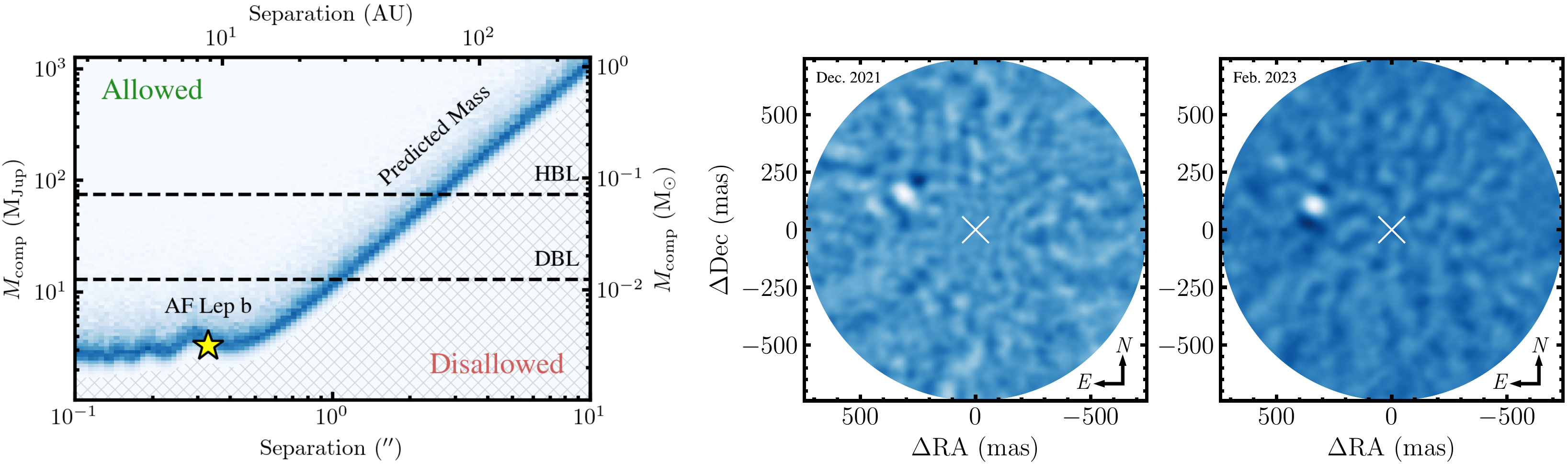}
\vskip -.1 in
    \caption{\emph{Left:} Predicted mass as a function of separation for AF~Lep~b from its HGCA astrometric acceleration. This plot is generated following \citet{Franson:2023bb}. The blue curve shows the masses and separations most consistent with the astrometric acceleration. Due to orbital phase, companions can occupy separations and masses above but not below the curve. The dynamical mass and semi-major axis of AF Lep b is denoted by the gold star. The planet is consistent with causing the acceleration. \emph{Right:} Keck/NIRC2 Imaging $L^\prime$ imaging of AF~Lep~b in December 2021 (left) and February 2023 (right). Here, S/N plots are displayed to enhance true sources over residual speckles. Note that the two images do not share the same colormap. We apply a Gaussian filter with a standard deviation of 1.5 pixels to average over pixel-to-pixel noise. AF Lep b is detected with a significance of $7.5\sigma$ in the December 2021 epoch and $13.0\sigma$ in the February 2023 epoch.
    \label{fig:images}}
\end{figure*}

AF Lep exhibits a significant\footnote{$\chi^2 = 77$, which corresponds to $8.5\sigma$ for two degrees of freedom.} astrometric acceleration between Hipparcos and Gaia DR3 in the HGCA. The HGCA provides three proper motion measurements: the proper motions at the Hipparcos and Gaia epochs and a joint average proper motion from the difference in sky position between the two missions. The average\footnote{The average acceleration is computed across a baseline of 12.6 years. Note, though, that the joint proper motion reflects the motion of AF Lep over the entire 25 years between the two missions.} tangential acceleration between the Gaia proper motion and the joint proper motion is \SI{0.0191 \pm 0.0024}{mas.yr^{-2}}, which corresponds to a physical acceleration of \SI{2.43 \pm 0.30}{m.s^{-1}.yr^{-1}} at the distance of AF Lep.
The predicted mass of a companion causing this acceleration would fall 
in the planetary regime if located within $\approx$30 au (see Figure \ref{fig:images}).  

\section{Observations \label{sec:observations}}

\subsection{Keck/NIRC2 Adaptive Optics Imaging\label{sec:nirc2}}

We obtained high-contrast imaging of AF Lep on UT 2021 December 21 and UT 2023 February 3 with the NIRC2 camera at W.M. Keck Observatory. Our observations were carried out in $L'$-band (3.426--\SI{4.126}{\micron}) with the Vector Vortex Coronagraph \citep[VVC;][]{serabynKeckObservatoryInfrared_2017} and natural guide star adaptive optics \citep{wizinowichAstronomicalScienceAdaptive_2013} with the Shack-Hartmann wavefront sensor. Imaging was taken in groups of 25 frames using the Quadrant Analysis of Coronagraphic Images for Tip-tilt Sensing \citep[\texttt{QACITS;}][]{hubyPostcoronagraphicTiptiltSensing_2015,hubyOnskyPerformanceQacits_2017} algorithm for centering the host star behind the VVC. Each sub-sequence includes an off-axis unsaturated image of the host star for flux calibration and sky background frames for both the science images and the flux calibration frame. The science exposures have integration times of \SI{0.18}{s} with 120 coadds. We read out a subarray of $512\times 512$ pixels for shorter read-out times. For the December 2021 epoch, our total integration time was 39.6 min, amounting to $35\fdg8$ of frame rotation. For the February 2023 epoch, our total integration time was 131.4 min, yielding $85\fdg5$ of frame rotation. The Differential Image Motion Monitor (DIMM) seeing during our December 2021 observations averaged  0$\farcs$5 arcsec; the average DIMM seeing during our February 2023 epoch was 0$\farcs$6 arcsec.

Our data reduction for both datasets begins with subtracting darks and flat-fielding. The \texttt{L.A.Cosmic} algorithm \citep{vandokkumCosmicRayRejection_2001} is then applied to identify and remove cosmic rays and hot pixels. We apply the distortion solution from \citet{serviceNewDistortionSolution_2016} to correct for geometric distortions in the optics of the imaging system. The sky background is modeled and subtracted from both the science and off-axis flux calibration frames using Principal Component Analysis (PCA) with the Vortex Image Processing \citep[\texttt{VIP;}][]{gomezgonzalezVipVortexImage_2017} package. Frames are co-registered to the sub-pixel level through a cross-correlation approach described in \citet{guizar-sicairosEfficientSubpixelImage_2008} and implemented in \texttt{scikit-image} \citep{vanderwaltScikitimageImageProcessing_2014} and \texttt{VIP}. Absolute centering is performed by fitting a negative Gaussian to the vortex core.

For carrying out PSF subtraction, we experimented with reductions from both \texttt{pyKLIP} \citep{wangPyklipPsfSubtraction_2015} and \texttt{VIP}. Both packages use Principal Component Analysis \citep[PCA;][]{soummerDetectionCharacterizationExoplanets_2012,amaraPynpointImageProcessing_2012} to model and subtract the host-star PSF. For the December 2021 dataset, the highest S/N is produced by a \texttt{pyKLIP} reduction with 20 annuli, 20 KL modes, and a value of one for the movement parameter. For the February 2023 imaging, the highest S/N is produced by a \texttt{VIP} reduction with 25 principal components. For the reduction of the February 2023 dataset, the PSF is subtracted in annuli, with frames only included in the reference PSFs for a given science frame if they have more than 1 FWHM of parallactic angle rotation for the annulus. For each sequence, we then compute S/N maps by measuring the flux in 0.5-FWHM-radius circular apertures. The noise level is estimated through the flux measured in non-overlapping circular apertures at the same separation. Figure \ref{fig:images} shows the S/N maps for our two epochs. We detect AF Lep b at a S/N of $7.5\sigma$ in the December 2021 imaging and $13.0\sigma$ in the February 2023 imaging.

To mitigate the introduction of systematics from the PSF-subtraction algorithm, we use the negative companion injection approach \citep[e.g.,][]{Marois:2008ei} to measure astrometry. A PSF template is generated by median-combining the off-axis flux calibration frames taken over each sequence. The inverse of this is then injected in the pre-processed frames at the approximate position, separation, and contrast of the companion and the PSF subtraction routine is performed. If the astrometry and photometry is well-matched to the companion, its signal will be removed in the post-processed images. Otherwise, the companion's parameters are adjusted, the negative template is re-injected, and the PSF subtraction routine is run again. 

The injected astrometry and photometry are first coarsely optimized using the \texttt{AMOEBA} downhill simplex algorithm \citep{nelderSimplexMethodFunction_1965}. We then use the \texttt{emcee} affine-invariant Markov-chain Monte Carlo (MCMC) ensemble sampler \citep{foreman-mackeyEmceeMcmcHammer_2013} with 100 walkers and $2 \times 10^4$ total steps to explore the astrometry and photometry parameter space. The sum of the residuals within a 4 FWHM-radius circular aperture of the source is used within the MCMC sampler to assess the match of the injected PSF to the companion. The first 30\% of each chain is discarded as burn-in. We evaluate convergence through both visual inspection of the chains and by verifying that subsections of the chains yield consistent astrometry. For the December 2021 dataset, the companion only appears at a high S/N in the PCA reductions when a parallactic angle cut is applied in constructing the reference PSFs for a given science image. We use 85 PCs for measuring the astrometry of the December 2021 epoch, which produces the highest S/N among the PCA reductions. For the February 2023 imaging, the companion appears at a high S/N in reductions without a parallactic angle cut, so we do not include this cut in the astrometry measurement to increase computational efficiency. When a parallactic angle cut is not performed, the maximum S/N is produced by 16 PCs, so we adopt this for the measurement. Ultimately, the number of PCs and PSF subtraction approach for the astrometry measurement and reduced images are each a balance between S/N, computational efficiency, and consistency in the approach for the two datasets.

The MCMC chains consist of separation $\rho$, position angle $\theta$, and a flux scaling factor from the PSF template. Following \citet{Franson:2022bl}, we combine the uncertainties on the astrometry from the MCMC run with the uncertainty in the distortion solution, north alignment, and plate scale. We also incorporate a \SI{4.5}{mas} centering uncertainty to account for the average performance of the \texttt{QACITS} algorithm \citep{hubyOnskyPerformanceQacits_2017}. Our final astrometry for the 2021 December epoch is $\rho = 338 \pm 11\, \mathrm{mas}$ and $\theta = 62\fdg8 \pm 1\fdg3$. Our astrometry for the 2023 February epoch is $\rho = 342 \pm 8\, \mathrm{mas}$ and $\theta = 72\fdg0 \pm 1\fdg0$. Converting the flux scaling factor to contrast yields $\Delta L' = 10.11 \pm 0.25\, \mathrm{mag}$ for the December 2021 epoch and $\Delta L' = 9.94 \pm 0.14\, \mathrm{mag}$ for the February 2023 epoch. Incorporating the $W1$ magnitude of AF Lep\footnote{Here we assume that $L' - W1 = 0$, since these filters are in the Rayleigh Jeans tail of the F8 host star's spectral energy distribution.} ($4.92 \pm 0.07 \, \mathrm{mag}$; \citealt{cutriVizierOnlineData_2012}), these correspond to apparent magnitudes of $L' = 15.03 \pm 0.26 \, \mathrm{mag}$ and $L' = 14.87 \pm 0.15 \, \mathrm{mag}$, respectively, and absolute magnitudes of $M_{L'} = 12.89 \pm 0.26 \, \mathrm{mag}$ (December 2021) and $M_{L'} = 12.72 \pm 0.15 \, \mathrm{mag}$ (February 2023).

\subsection{Transiting Exoplanet Survey Satellite Light Curve\label{sec:tess}}

AF Lep was observed with the Transiting Exoplanet Survey Satellite \citep[TESS;][]{Ricker2015} in Sectors 5, 6, and 32, and processed following the procedure described in \citet{Bowler2023}. The Science Processing Operations Center (SPOC) Pre-search Data Conditioning Simple Aperture Photometry \citep[PDCSAP;][]{Jenkins2016} light curve was downloaded using the \texttt{lightkurve} package \citep{Lightkurve2018}. Outlier photometric points are removed by flattening the light curve with a high-pass Savitzky-Golay filter and identifying all data outside of three standard deviations. 

Clear modulations are evident in the final light curve with an amplitude of 1--2\%. The rotation period and uncertainty measurements are calculated by producing a Generalized Lomb-Scargle periodogram \citep{Zechmeister2009}, fitting a Gaussian to the highest periodogram peak, and adopting the resulting mean and standard deviation. This yields a raw periodicity of 1.007 $\pm$ 0.007 d.  
However, starspots on AF Lep could be located at mid-latitudes rather than the equator.  If differential rotation is present then this will impact the interpretation of these modulations as a rotation period.  To account for this, we follow \citet{Bowler2023} by inflating the uncertainty assuming a solar-like absolute shear of 0.07 rad d$^{-1}$.  Adding this uncertainty from potential differential rotation in quadrature with the uncertainty from the periodogram analysis gives a final adopted value for the rotation period of 1.007 $\pm$ 0.009 d.  This is similar to values of 0.966 $\pm$ 0.002~d from \citet{Messina:2017cx}, 1.03 $\pm$ 0.03~d from \citet{gaiacollaborationVizierOnlineData_2022}, and 1.01 $\pm$ 0.05 d from \citet{ZunigaFernandez:2021if}\footnote{Note that the \citet{ZunigaFernandez:2021if} measurement only used Sector 5 of the TESS lightcurve of AF Lep, which may contribute to its higher uncertainty.}.

\begin{figure}

        \centering
        \includegraphics[width=\linewidth]{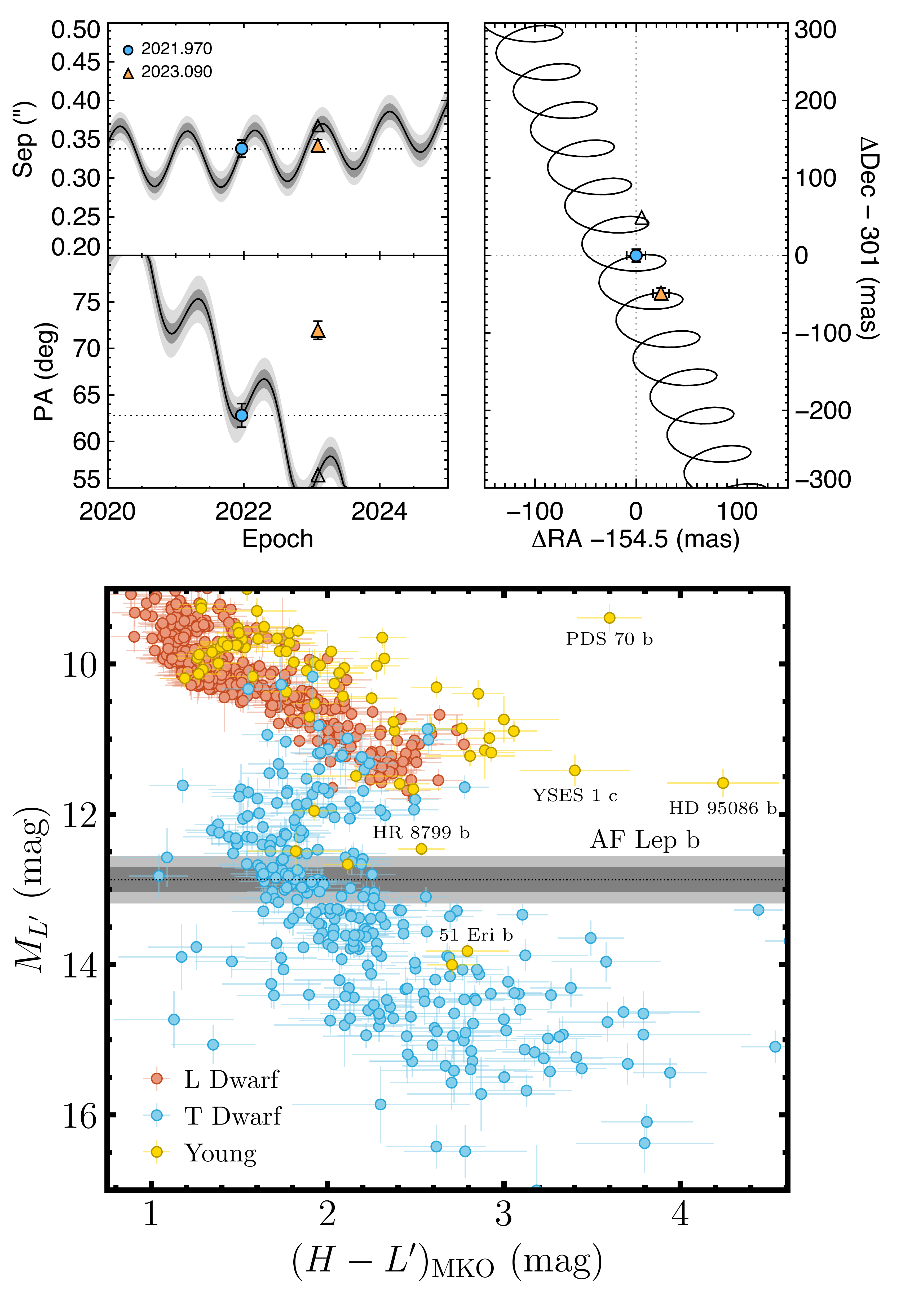}
        \caption{\emph{Top:} Test for common proper motion based on the two epochs of relative astrometry between AF Lep A and b. The left panels show the predicted relative motion of a stationary source in separation (top) and P.A. (bottom) with respect to the first epoch of relative astrometry (blue circles).  Gray shaded regions represent 1- and 2-$\sigma$ uncertainties.  The predicted position at 2023.090 is plotted as an open triangle, and the orange triangles are our measurements.  Our 2023 observations significantly disagree with a stationary background source model at the $11\sigma$-level in P.A.  The right panel shows the same comparison, but in $\Delta$RA and $\Delta$Decl. instead of separation and P.A. \emph{Bottom:} Comparison of our $L^\prime$ photometry of AF Lep b to the $M_{L^\prime}$--$(H-L^\prime)$ CMD. The grey regions show the 1- and 2-$\sigma$ uncertainties on the $L'$ magnitude from our 2023 imaging. The background points show L (red) and T (blue) dwarfs from The UltracoolSheet. Sources marked as young or low surface gravity are plotted in yellow. Young objects in the compilation are sources with independent age constraints ${<}300 \, \mathrm{Myr}$. Low surface gravity sources have spectroscopic features of youth based on near-infrared spectroscopy. We also add substellar companions with $H$ and $L^\prime$ photometry compiled within the \texttt{species} package \citep{stolkerMiraclesAtmosphericCharacterization_2020}. The $L'$ photometry of AF Lep b implies an early-T dwarf spectral type.
        \label{fig:cpm_cmd_masspred}}
\end{figure}

\begin{figure*}
        \centering
        \includegraphics[width=0.9\textwidth]{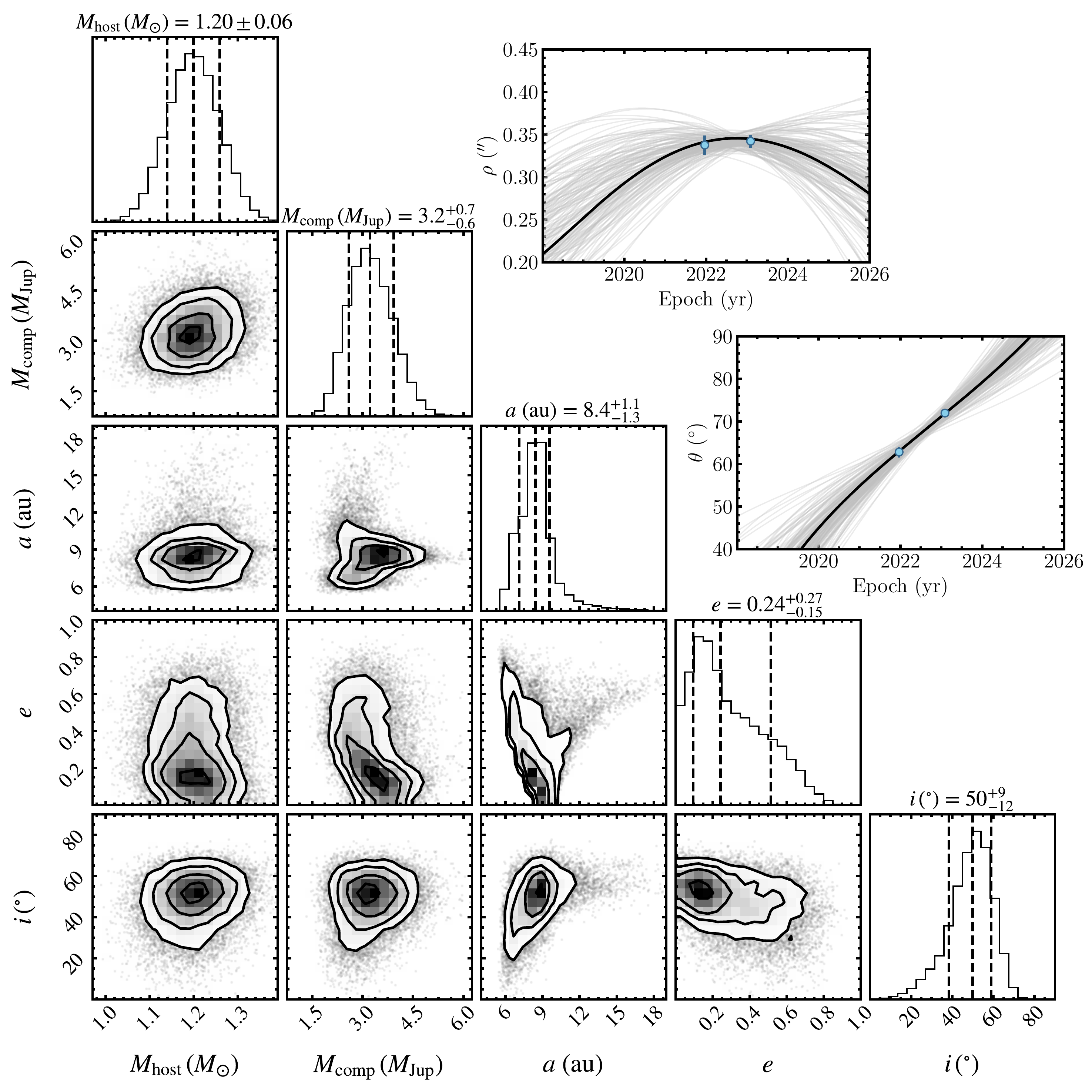}
    \vskip -.1 in
        \caption{Joint posterior distributions of $M_{\mathrm{host}}$, $M_{\mathrm{comp}}$, $a$, $e$ and $i$ for the orbit fit of AF Lep b. The diagonal panels show the marginalized distribution for each parameter. The contours on off-diagonal panels denote $1\sigma$, $2\sigma$, $3\sigma$, and $4\sigma$ levels in the joint posteriors. The upper right plots compare the separation and position angle of AF Lep b over time against a swarm of orbits drawn from the orbit fit, shown in grey. The black curves highlight the maximum-likelihood orbit.
        \label{fig:orbitfit}}
    \end{figure*}

\section{Results \label{sec:results}}

\subsection{Common Proper Motion \label{sec:cpm}}

Our second epoch of relative astrometry from February 2023 shows no significant change in separation compared to the first epoch taken 13 months earlier (338 $\pm$ 11~mas compared to 342 $\pm$ 8~mas in 2023), but a large counter-clockwise change in P.A. (62.8 $\pm$ 1.3$\degr$ compared to 72.0 $\pm$ 1.0$\degr$ in 2023). To test whether this is consistent with a stationary background source, perfect comovement with the host star, or comovement plus orbital motion, we compare our measurements with the predicted trajectories in Figure~\ref{fig:cpm_cmd_masspred} using the host star sky coordinates, parallax, and proper motion.
At epoch 2023.090, the predicted separation for a stationary source is 367 $\pm$ 10~mas and the predicted P.A. is 56.4 $\pm$ 1.0$\degr$.  
Although the second epoch separation is similar, our measured P.A. of AF Lep b differs by about $9\fdg2$ between the two observations.  This corresponds to a 2.0-$\sigma$ discrepancy in separation and an 11.0-$\sigma$ discrepancy in P.A.
We thus find that the common proper motion test strongly favors the source being a gravitationally bound companion rather than a background star. The apparent motion between the two epochs is therefore a result of orbital motion, not seen in separation (0.3$\sigma$), but detected at the 5.6$\sigma$-level in P.A.  Assuming linear motion, the P.A. of AF Lep b is increasing by about 8$\degr$ yr$^{-1}$.

\subsection{Color Magnitude Diagram and Inferred Mass\label{sec:cmd}}
We generate a $M_{L^\prime}$--$(H-L^\prime$) color magnitude diagram (CMD) of nearby substellar sources to compare against the $L'$ photometry of AF Lep b using The UltracoolSheet \citep{bestwilliamm.j.UltracoolsheetPhotometryAstrometry_2020}. Single objects in the compilation with $W1$-band magnitudes, parallaxes, and $H$-band magnitudes in the Maunakea Observatories (MKO) filter system \citep{tokunagaMaunaKeaObservatories_2005} are selected. We convert the $W1$-band magnitudes to $L'$-band magnitudes using the $M_{W1}$--$L'$ relation derived in \citet{Franson:2023bb}. For the absolute $L'$ magnitude of AF Lep b, we adopt the value from our 2023 February imaging of $M_L' = 12.72 \pm 0.15 \, \mathrm{mag}$. Our CMD is shown in the lower panel of Figure \ref{fig:cpm_cmd_masspred}. The photometry of AF Lep b would imply an early T-dwarf spectral type, although at the young age and low gravity of the companion, thick clouds may significantly redden the spectrum (e.g., \citealt{Liu:2016co}; \citealt{Faherty:2016fx}).

From our absolute $L'$ photometry, we can determine the model-inferred mass of AF Lep b given its age of $24 \pm 3$ Myr. Our approach is to draw from the age and absolute $L'$ magnitude distributions $10^6$ times and interpolate a given hot-start model grid to determine the corresponding mass distribution. We perform this procedure for both Cond \citep{baraffeEvolutionaryModelsCool_2003} and the \texttt{ATMO-2020} \citep{phillipsNewSetAtmosphere_2020} models. This yields hot-start model-inferred masses of $4.53^{+0.32}_{-0.35} \, M_\mathrm{Jup}$ for Cond and $5.4 \pm 0.5 \, M_\mathrm{Jup}$ for \texttt{ATMO-2020}. We can also infer masses from the \citet{burrowsNongrayTheoryExtrasolar_1997} and \citet{saumonEvolutionDwarfsColor_2008} grids by determining the bolometric luminosity of AF Lep b and interpolating the models. We base our bolometric correction on HR 8799 b, which has a similar $L'$ absolute magnitude to AF Lep b ($M_{L'} = 12.66 \pm 0.11 \, \mathrm{mag}$; \citealt{Marois:2010gp}). Following the procedure outlined in \citet{brandtFirstDynamicalMass_2021}, the bolometric lumnosity of HR 8799 b is $\log(L/L_{\mathrm{bol}}) = -4.79 \pm 0.10 \, \mathrm{dex}$, which corresponds to a bolometric correction $BC(L') = -0.68 \pm 0.27 \, \mathrm{mag}$. Applying this correction, the bolometric luminsoity of AF Lep b is then $\log(L/L_{\mathrm{bol}}) = -4.81 \pm 0.13 \, \mathrm{dex}$. This produces inferred masses of $6.2^{+1.0}_{-0.9} \, M_\mathrm{Jup}$ for \citet{burrowsNongrayTheoryExtrasolar_1997} and $5.6^{+0.7}_{-0.7} \, M_\mathrm{Jup}$ for the \citet{saumonEvolutionDwarfsColor_2008} grid with the hybrid prescription for clouds. The cloud-free and cloudy versions of the \citet{saumonEvolutionDwarfsColor_2008} models yield masses of $6.1^{+0.9}_{-0.8} \, M_\mathrm{Jup}$ and $6.3^{+1.0}_{-0.9} \, M_\mathrm{Jup}$, respectively.

\begin{deluxetable*}{lccc} 
    \tablecaption{\label{tab:elements}AF Lep b Orbit Fit Results}
    \tablehead{\colhead{Parameter} & \colhead{Median $\pm 1\sigma$} & \colhead{95.4\% C.I.} & \colhead{Prior}}
    \startdata
    \multicolumn{4}{c}{Fitted Parameters} \\
    \hline
    $M_{\mathrm{comp}}$ $(M_\mathrm{Jup})$ & ${3.2}_{-0.6}^{+0.7}$ & (2.1, 4.7) & $1/M_{\mathrm{comp}}$ (log-flat), $M_\mathrm{comp} \in (0, 10^3 \, \si{M_\odot}]$\tablenotemark{c}\\
    $M_{\mathrm{host}}$ $(M_\mathrm{\odot})$ & $1.20 \pm 0.06$ & (1.08, 1.32) & $\SI{1.20 \pm 0.06}{M_\odot}$ (Gaussian)\\
    $a$ $(\mathrm{au})$ & ${8.4}_{-1.3}^{+1.1}$ & (6.2, 12.8) & $1/a$ (log-flat), $a \in (0, 2 \times 10^5 \, \si{au}]$\tablenotemark{c}\\
    $i_o$ $(\si{\degree})$ & ${50}_{-12}^{+9}$ & (23, 66) & $\sin (i)$, $\SI{0}{\degree} < i < \SI{180}{\degree}$\\
    $\sqrt{e} \sin{\omega}$ & ${-0.1}_{-0.4}^{+0.5}$ & (-0.8, 0.7) & Uniform, $-\sqrt{e} \leq \sqrt{e} \sin \omega \leq \sqrt{e}$\\
    $\sqrt{e} \cos{\omega}$ & ${-0.16}_{-0.26}^{+0.40}$ & (-0.60, 0.58) & Uniform, $-\sqrt{e} \leq \sqrt{e} \cos \omega \leq \sqrt{e}$\\
    $\Omega$ $(\si{\degree})$ & ${66}_{-22}^{+13}$\tablenotemark{a} & (10, 122)\tablenotemark{a} & Uniform, $\SI{-180}{\degree} \leq \omega \leq \SI{540}{\degree}$\\
    $\lambda_{\mathrm{ref}}$ $(\si{\degree})$\tablenotemark{b} & ${160}_{-90}^{+70}$ & (10, 350) & Uniform, $\SI{-180}{\degree} \leq \lambda_{\mathrm{ref}} \leq \SI{540}{\degree}$\\
    Parallax $(\si{mas})$ & $37.254 \pm 0.020$ & (37.214, 37.293) & $\SI{37.254 \pm 0.020}{mas}$ (Gaussian)\\
    $\mu_\alpha$ ($\si{mas.yr^{-1}}$) & $17.111 \pm 0.024$ & (17.074, 17.154) & Uniform, $\mu_\alpha \in (-\infty, \infty)$\tablenotemark{c}\\
    $\mu_\delta$ ($\si{mas.yr^{-1}}$) & ${-49.183}_{-0.017}^{+0.021}$ & (-49.211, -49.142) & Uniform, $\mu_\delta \in (-\infty, \infty)$\tablenotemark{c}\\
    RV Jitter $\sigma_{\mathrm{RV}}$ ($\si{m.s^{-1}}$) & ${177}_{-30}^{+39}$ & (124, 265) & $1/\sigma_{\mathrm{RV}}$ (log-flat), $\sigma_{\mathrm{RV}} \in (0, 10^3 \, \si{m.s^{-1}}]$\tablenotemark{c}\\
    \hline
    \multicolumn{4}{c}{Derived Parameters} \\
    \hline
    $P$ (yr) & $22 \pm 5$ & (14, 42) & . . .\\
    $e$ & ${0.24}_{-0.15}^{+0.27}$ & (0.02, 0.71) & . . .\\
    $\omega$ $(\si{\degree})$ & ${210}_{-100}^{+80}$ & (20, 340) & . . .\\
    $T_0$ $(\mathrm{JD})$ & ${2456700}_{-1200}^{+800}$ & (2448400, 2458500) & . . .\\
    $T_0$ (yr) & ${2014.1}_{-3.2}^{+2.1}$ & (1991.3, 2019.1) & . . .\\
    $q$ $(=M_{\mathrm{comp}}/M_{\mathrm{host}})$ & $0.0026 \pm 0.0005$ & (0.0017, 0.0037) & . . .
    \enddata
    \tablenotetext{a}{The posterior distribution for $\Omega$ consists of two distinct peaks separated by \SI{180}{\degree}. The values shown in the table correspond to the lower peak. The other peak is located at ${250}_{-40}^{+60}{}^\circ$ with a 95.4\% confidence interval of (170, 350).}\tablenotetext{b}{Mean longitude at the reference epoch of 2010.0.}\tablenotetext{c}{Note that these are improper priors, since they cannot be normalized to one.}
    
    \end{deluxetable*}

\subsection{Orbit Fit and Dynamical Mass \label{sec:orbit}}
Here, we perform a joint orbit fit with \texttt{orvara} \citep{brandtOrvaraEfficientCode_2021} of our Keck/NIRC2 relative astrometry, archival radial velocities (RVs), and the HGCA proper motions. The archival RVs were published in \citet{butlerLcesHiresKeck_2017} and consist of 20 measurements with the HIRES spectrograph \citep{vogtHiresHighresolutionEchelle_1994} on Keck I. The observations were taken from September 2002 to November 2013 and have a median reported uncertainty of \SI{63.2}{m.s^{-1}} and an rms of \SI{188.7}{m.s^{-1}}.

\texttt{orvara} uses the parallel-tempered MCMC (PT-MCMC) ensemble sampler in \texttt{emcee} to infer the orbit element posteriors. The following orbital elements are directly fit: companion mass $M_{\mathrm{comp}}$, primary mass $M_{\mathrm{host}}$, semi-major axis $a$, inclination $i_o$, longitude of ascending node $\Omega$, longitude at the reference epoch of 2010.0 $\lambda_{\mathrm{ref}}$, and RV jitter $\sigma_{\mathrm{RV}}$. Eccentricity $e$ and argument of periastron $\omega$ are parameterized as $\sqrt{e} \sin \omega$ and $\sqrt{e} \cos \omega$ to avoid the Lucy-Sweeney bias against circular orbits \citep{lucySpectroscopicBinariesCircular_1971}. \texttt{orvara} analytically marginalizes over parallax, barycentric proper motion $\mu_{\alpha}$ and $\mu_{\delta}$, and RV instrumental zeropoints. We adopt uniformative priors for all quantities except host-star mass, where we take a prior of $1.2 \pm 0.06\, \mathrm{M_\odot}$ from its mass in \citet{Kervella:2022aa}. We use 20 temperatures, 100 walkers, and $5 \times 10^5$ total steps (5000 steps per walker) in the orbit fit. The first 20\% of each chain is discarded as burn in. Convergence is assessed by verifying that multiple runs with different starting parameters produce consistent parameter posteriors.

Table \ref{tab:elements} shows the results of our orbit fit. Figure \ref{fig:orbitfit} displays the posterior distributions for selected orbit elements alongside a comparison of the relative astrometry of AF Lep b to a swarm of orbits drawn from the fit. We measure a dynamical mass of $3.2^{+0.7}_{-0.6}\, M_\mathrm{Jup}$, semi-major axis of $8.4^{+1.1}_{-1.3} \, \mathrm{au}$, orbital inclination of $50^{+9}_{-12} {}^\circ$, and eccentricity of $0.24^{+0.27}_{-0.15}$. The orbital period is $22 \pm 5$ yr and the time of periastron is $T_0 = 2014.1^{+2.1}_{-3.2}$. AF Lep b and its host star have a mass ratio of $0.0026 \pm 0.0005$.

To assess the potential of systematics in the relative astrometry or Hipparcos-Gaia proper motions impacting the companion's dynamical mass, we perform a suite of additional orbit fits to synthetic data with systematic offsets from the nominal measurements at the $2$- and $4$-$\sigma$ levels. Each of these fits uses 20 temperatures, 100 walkers, and $10^5$ steps. We first vary each epoch of the relative astrometry by $\pm2\sigma$ and $\pm4\sigma$ while keeping the other epoch constant. Here, the separation and position angle are both offset by the same amount for each instance (for example, $+4\sigma$ in separation and $+4\sigma$ in position angle). We find that the dynamical mass is resilient to systematics at the $2$- and $4$-$\sigma$ levels for both epochs, with all resultant companion masses being consistent with our dynamical mass to within $1\sigma$. The companion mass posteriors from the mock runs have similar uncertainties to the dynamical mass. Changes to the HGCA proper motions have a larger impact on the dynamical mass. Shifting the Hipparcos-Gaia proper motion by ${-}2\sigma$ or the Gaia proper motion by ${+}2\sigma$ decreases the dynamical mass by about $1\sigma$ from our nominal case to ${\approx} 2.5 \, M_\mathrm{Jup}$. Applying the opposite offset increases the dynamical mass by about $1\sigma$ to ${\approx}3.9 \, M_\mathrm{Jup}$. Offsets at the $4$-$\sigma$ level change the dynamical mass by about $2\sigma$ to about $1.7 \, M_{\mathrm{Jup}}$ or $4.5 M_\mathrm{Jup}$. We thus find that the dynamical mass is more significantly affected by systematics in the Hipparcos-Gaia proper motions than the relative astrometry.

Our dynamical mass measurement is slightly lower than model-inferred masses from the planet's $L'$ photometry and associated bolometric luminosity. It is $1.7\sigma$ lower than the Cond-inferred mass of $4.53^{+0.32}_{-0.35} \, M_\mathrm{Jup}$, $2.6\sigma$ lower than the mass inferred via \texttt{ATMO-2020} of $5.4 \pm 0.5 \, M_\mathrm{Jup}$, $2.7\sigma$ lower than the mass inferred from \citet{burrowsNongrayTheoryExtrasolar_1997} of $6.2^{+1.0}_{-0.9} \, M_\mathrm{Jup}$, and $2.4\sigma$ lower than the \citet{saumonEvolutionDwarfsColor_2008}-inferred mass with the hybrid prescription for clouds. Here, one-sided Gaussian-equivalent $\sigma$ values are determined following \citet{Franson:2023bb}. All inferred masses are generated using hot-start models, which treat planets as having arbitrarily large radii and high entropies at early ages as an initial condition. Cold- \citep[e.g.,][]{marleyLuminosityYoungJupiters_2007,fortneySyntheticSpectraColors_2008} and warm-start \citep[e.g.,][]{spiegelSpectralPhotometricDiagnostics_2012,marleauConstrainingInitialEntropy_2014} models start with lower amounts of initial entropy to emulate the loss of energy through accretion shocks in planet formation. Since this lowers the luminosity of planets at young ages relative to hot-start models, the discrepancy between our dynamical mass and model predictions is \emph{larger} if a cold- or warm-start model is assumed. 

One way this modest discrepancy can be reconciled is if the planet is younger than the star, perhaps due to delayed formation in a disk. This would cause the planet to be brighter than expected based on the host star age. For the case of AF Lep b, the planet would require an age of $16^{+7}_{-5}$ Myr for the Cond mass to match the dynamical mass. Other models require lower masses; for the prediction from the hybrid \citet{saumonEvolutionDwarfsColor_2008} model to match the dynamical mass, the planet would need to be $9^{+4}_{-3}$ Myr. This would imply a formation timescale of $\approx5{-}15$ Myr, which would suggest formation via core accretion, which takes many Myr to operate, over rapid giant planet formation via disk instability. Typical protoplanetary disk lifetimes are 5--10~Myr for Sun-like and lower-mass (${<}2 \, M_\odot$) stars \citep{pfalznerMostPlanetsMight_2022}. Other possibilities for resolving this discrepancy include a younger age of the system, young L/T transition objects being more luminous than predicted by current evolutionary models, and another planet in the system biasing the dynamical mass measurement.

\subsection{Planet-Disk Interactions}
There are several potential avenues for an inner planet to dynamically interact with an outer debris disk. One possibility is the planet truncating the disk's inner edge. Due to the close separation of AF Lep b from its host star, it is unlikely that the planet is truncating the disk from its current orbit. \citet{Pearce:2022aa} found that the disk inner edge could be truncated by a $1.1 \pm 0.2 \, M_\mathrm{Jup}$ planet at a semi-major axis of $35 \pm 6 \, \mathrm{au}$. Larger objects at closer-in semi-major axes could also produce the same effect, but the dependence between mass and semi-major axis is sufficiently steep that by the $8 \, \mathrm{au}$ semi-major axis of AF Lep b, only stellar companions are capable of truncating the disk (see Equation 6 of \citealt{Pearce:2022aa}). This remains the case if we allow for the possible eccentricity of AF Lep b and the uncertain location of the unresolved disk. If the debris disk was truncated by planets, then either additional, unseen planets are responsible or AF~Lep~b was closer to the disk in the past and has since migrated inwards.

While AF~Lep~b is unlikely to be truncating the disk, its possible eccentricity could drive the disk into an asymmetric shape. Equation 5 in \cite{Pearce:2022aa} relates the semi-major axis and eccentricity of a planet's orbit to the eccentricities of a debris disk's edges. A planet at ${8 \; \rm au}$ with an eccentricity of 0.25 would drive a narrow disk at ${46 \; \rm au}$ to an eccentricity of $\sim0.1$. The disk is currently unresolved, so no eccentricity or width information is available. Future resolved observations of the disk have the potential to determine the disk's eccentricity and directly examine the interaction between an eccentric planet and its debris disk.

A third possibility is stirring, where the planet excites planetesimals in the disk, causing them to collide and release the observed dust. Equation 15 of \citet{mustill2009} gives the mass of an interior planet required to stir debris as a function of planet eccentricity and location. For an eccentricity of 0.25 and semi-major axis of $8\, \mathrm{au}$, planets with masses greater than $0.45 \, M_\mathrm{Jup}$ are able to stir the AF Lep debris disk. AF Lep b, with its mass of $3.2^{+0.7}_{-0.6} \, M_\mathrm{Jup}$, may be actively stirring the disk. This is significant, as there is growing evidence that many debris disks are unable to stir themselves via self gravity (\citealt{Krivov2018,Pearce:2022aa}, though see also \citealt{najita2022}). Planetary stirring could therefore be prevalent in generating the observed dust in debris disks. Note that if AF Lep b has a small but non-zero eccentricity, it could still stir the disk, provided the eccentricity is above ${\approx}0.04$. This threshold is determined by rearranging Equation 15 of \citet{mustill2009} to give the minimum eccentricity capable of stirring the disk as a function of planet mass.

\subsection{Stellar Inclination and Minimum Obliquity \label{sec:inclination}}

The line-of-sight inclination of AF Lep, $i_*$, can be inferred from its rotation period, stellar radius, and projected rotational velocity.  We follow the Bayesian framework of \citet{Masuda:2020dp} to derive $i_*$ using the analytical expression for its posterior distribution from \citet[][, their Equation 9]{Bowler2023}, which accounts for the correlation between projected and equatorial rotational velocities.  
Here we adopt the \emph{TESS} rotation period of 1.007 $\pm$ 0.009 d, the $v \sin i_*$ value of 50 $\pm$ 5 km s$^{-1}$ from \citet{Glebocki:2005aa}, and radius of 1.25 $\pm$ 0.06 $R_{\odot}$ from \citet{Kervella:2004aa}. The Maximum a posteriori (MAP) value is 54$\degr$. The 68\% credible interval spans 45--65$\degr$ and
the 95\% credible interval spans 39--82$\degr$.

The stellar inclination together with the inclination 
of the orbital plane of AF Lep b ($i_o$) provides information about
the minimum stellar obliquity---whether the spin axis of 
AF Lep is aligned or misaligned with respect to the planet.
Our orbit fit indicates an inclination of $i_o$ = 50$^{+9}_{-12}\degr$. This is in excellent agreement with the stellar inclination of $i_*$ = 54$^{+11}_{-9}\degr$.  The polar position angle of AF Lep---which describes its absolute orientation in the sky plane---is
not known and can therefore not be compared with the longitude of ascending node of the orbit to assess the true obliquity angle.  However, the agreement of the inclination angles is nevertheless consistent with spin-orbit alignment, which reinforces trends seen in other systems with imaged planets but differs from the more diverse architectures of brown dwarf companions (\citealt{Bowler2023}).  The debris disk around AF Lep has not been resolved; if future observations resolve the disk, its inclination could be compared with the stellar and orbital inclinations to further assess angular momentum alignment in the AF Lep system.

\section{Summary \label{sec:summary}}

AF Lep b is a young giant planet directly imaged inside the debris disk of a Sun-like star in the 24 $\pm$ 3~Myr $\beta$~Pic moving group.  The system was identified as a promising target based on the low-amplitude (\SI{2.43 \pm 0.30}{m.s^{-1}. yr^{-1}}) astrometric acceleration the planet induces on its host star.  Our main conclusions are summarized below:

\begin{itemize}
\item Our two epochs of NIRC2 observations in $L'$ with the VVC separated by 13 months confidently show that AF Lep b is gravitationally bound and is undergoing orbital motion around its host star.  We can rule out AF Lep b being a background star at the 11-$\sigma$ level and orbital motion is detected at the 6-$\sigma$ level.

\item AF Lep b has a contrast of $\Delta L' = 9.94 \pm 0.14 \, \mathrm{mag}$ at 0$\farcs$34 and an absolute magnitude of $M_{L'}$ = 12.72 $\pm$ 0.15~mag. Based on the age of the host star, and assuming the planet was formed at the same time, 
the implied planet mass from hot-start evolutionary models is ${\approx} 4{-}6 \, M_\mathrm{Jup}$.

\item Our orbit fit incorporating the relative astrometry, astrometric
acceleration, and radial velocities yields a dynamical mass of 3.2$^{+0.7}_{-0.6}$~$M_\mathrm{Jup}$, a semi-major axis of 8.4$^{+1.1}_{-1.3}$~au, and a modest eccentricity of 0.24$^{+0.27}_{-0.15}$ for AF Lep b. The mass of AF Lep b is too low at an orbital distance of $\approx$8~au to truncate the outer debris disk, which is located at $\sim$46 $\pm$ 9~au (\citealt{Pearce:2022aa}).  Another massive planet or series of low-mass planets may be present at wider separations.

\item The dynamical mass (3.2$^{+0.7}_{-0.6}$~$M_\mathrm{Jup}$) is somewhat lower than predicted masses (${\approx}4{-}6 \, M_\mathrm{Jup}$) from hot-start models assuming a nominal age of 24 $\pm$ 3~Myr from the host star. This could be caused by a slightly younger age for the host star, delayed formation in a disk, systematic uncertainties in the evolutionary models, or another companion in the system.

\item  The stellar inclination ($i_*$ = 54$^{+11}_{-9}\degr$) and orbital inclination ($i_o$ = 50$^{+9}_{-12}\degr$) are in good agreement, which is consistent with the system having spin-orbit alignment, although polar position angle of the star is unknown so this agreement represents a lower limit on the true obliquity.

\end{itemize}

AF Lep joins other young planet hosts with debris disks such as $\beta$ Pic, HR 8799, HD 206893, and HD 95086, reinforcing indications of a higher frequency of long-period planets orbiting stars hosting debris disks (\citealt{Meshkat:2017jka}).  Within the $\beta$ Pic moving group, the mass of AF Lep b is higher than 51 Eri b  ($\approx$2~$M_\mathrm{Jup}$; \citealt{Macintosh:2015fw}) and lower than $\beta$ Pic b and c (8--9~$M_\mathrm{Jup}$; \citealt{Nowak:2020cc}), contributing to a growing sequence of giant planets with common ages. AF Lep b is the lowest-mass imaged planet with a dynamical mass. We expect it to become an excellent target for follow-up orbital and atmospheric characterization.

\section{Acknowledgements}
K.F. acknowledges support from the National Science Foundation Graduate Research Fellowship Program under Grant No. DGE 2137420.
B.P.B. acknowledges support from the National Science Foundation grant AST-1909209, NASA Exoplanet Research Program grant 20-XRP20$\_$2-0119, and the Alfred P. Sloan Foundation.
Y.Z. acknowledges support from the Heising-Simons Foundation 51 Pegasi b Fellowship. T.D.P. is supported by Deutsche Forschungsgemeinschaft (DFG) grants KR 2164/14-2 and KR 2164/15-2. Q.H.T. and B.P.B. acknowledge the support from a NASA FINESST grant (80NSSC20K1554). This work was supported by a NASA Keck PI Data Award, administered by the NASA Exoplanet Science Institute. 
This research has made use of the VizieR catalogue access tool, CDS, Strasbourg, France (DOI: 10.26093/cds/vizier). The original description of the VizieR service was published in 2000, A\&AS 143, 23. This publication makes use of data products from the Wide-field Infrared Survey Explorer, which is a joint project of the University of California, Los Angeles, and the Jet Propulsion Laboratory/California Institute of Technology, funded by the National Aeronautics and Space Administration. This work has benefited from The UltracoolSheet at \href{http://bit.ly/UltracoolSheet}{http://bit.ly/UltracoolSheet}, maintained by Will Best, Trent Dupuy, Michael Liu, Rob Siverd, and Zhoujian Zhang, and developed from compilations by \citet{dupuyHawaiiInfraredParallax_2012}, \citet{dupuyDistancesLuminositiesTemperatures_2013}, \citet{liuHawaiiInfraredParallax_2016}, \citet{bestPhotometryProperMotions_2018}, \citet{bestVolumelimitedSampleL0t8_2021}.

The authors wish to recognize and acknowledge the very significant cultural role and reverence that the summit of Maunakea has always had within the indigenous Hawaiian community. We are most fortunate to have the opportunity to conduct observations from this mountain.

\facilities{Keck:II (NIRC2)}
\software{\texttt{VIP} \citep{gomezgonzalezVipVortexImage_2017}, \texttt{pyKLIP} \citep{wangPyklipPsfSubtraction_2015}, \texttt{orvara} \citep{brandtOrvaraEfficientCode_2021}, \texttt{ccdproc} \citep{craigAstropyCcdprocV1_2017}, \texttt{photutils} \citep{bradleyAstropyPhotutilsV0_2019}, \texttt{astropy} \citep{astropycollaborationAstropyCommunityPython_2013,astropycollaborationAstropyProjectBuilding_2018}, \texttt{pandas} \citep{mckinneyDataStructuresStatistical_2010}, \texttt{matplotlib} \citep{hunterMatplotlib2dGraphics_2007}, \texttt{numpy} \citep{harrisArrayProgrammingNumpy_2020}, \texttt{scipy} \citep{virtanenScipyFundamentalAlgorithms_2020}, \texttt{emcee} \citep{foreman-mackeyEmceeMcmcHammer_2013}, \texttt{corner} \citep{foreman-mackeyCornerPyScatterplot_2016}, \texttt{lightkurve} \citep{lightkurvecollaborationLightkurveKeplerTess_2018}, \texttt{scikit-image} \citep{vanderwaltScikitimageImageProcessing_2014}}

\bibliography{references}{}
\bibliographystyle{aasjournal}
\newpage

\end{document}